\documentstyle[12pt]{article}

\begin{document}
\title{{\bf Nilpotent Marsh and SUSY QM}}

\author{{\bf V. P. Akulov}\\
{\sl City College of City University of New York,}\\
{\sl 138th St. Convent Ave , New York, NY 10031, USA} \\ and
\\{\bf Steven Duplij}\thanks{E-mail:
{\tt Steven.A.Duplij@univer.kharkov.ua}} \thanks{
Internet: http://{\tt www-home.univer.kharkov.ua/\~{}duplij}}\\
{\sl Theory Group, Nuclear Physics Laboratory, }\\{\sl Kharkov State
University, Kharkov 310077, Ukraine}}

\maketitle

\begin{abstract}
We consider the nilpotent additions to classical trajectories in
supersymmetric and nonsupersymmetric theories. The condition of
anilpotence of action on some generalized solutions leads to the Witten
supersymmetric Lagrangian. The condition of anilpotence of topological
charge is the same as one of superpotential with spontaneous broken
supersymmetry. We should vanish half of Grassmann constants of integration,
because in this case only we obtain the same number of normalized bosonic
and fermionic zero modes.

\end{abstract}
\newpage
\section{Introduction}

This paper is dedicated to Victor Isaakovich Ogievetsky who was an
outstanding scientist and brilliant man. He gave a great contribution to
quantum field theory, supersymmetry and supergravity.

In recent time the interest to extended supersymmetric quantum mechanics
\cite{wit2} with N supersymmetries \cite
{and/can/ded/iof,dia/ent,man0,low,pab/set/ste1,pab/set/ste2} is greatly
renewed. This is connected on one side with dimensional reduction of
additional dimensions of D-brane where one can receive SQM limit, which is
more simple object. Maldacena's correspondence \cite{mald} between string
theory on AdS spaces and superconformal quantum theories attracts attention
to the structure of extremal black holes near the horizon which is described
by the supersymmetric quantum mechanics \cite{cla/der/kal}.

On the other side
the supermembrane in light cone gauge gives rise to SQM system with SU(N)
gauge symmetry \cite{dew/mar/nic,low}. The D(-1) branes or D-instantons have the
world volume which is just a point; they are topological objects and carry
out topological charge.

Instantons as such solitons have zero modes connected with derivative by
constants of integration. It is well known that these solutions (as a rule)
break one half supersymmetries. For any supersymmetric theory the number of
bosonic zero modes should be the same as fermionic ones.

We want to consider a full class of solutions for classical equations of
motion, which consist of both bosonic and Grassmann constants of integration
\cite{aku,aku/vol2,aku/dup,aku/pas,man0}. Fields take their values in
Grassmann algebra with an even part and an odd part. But the action have to
belong only to the even part and should not have any nilpotent addition,
because the action is connected with a measure defined by functional
integration in the path integral approach.

We here concentrate our attention mainly on the conditions which give
absence of nilpotent additions to the action on the generalized classical
trajectories and receive the connection between bosonic and fermionic parts
of the interaction.

The plan of the paper is the following. In Section 2 we introduce the
superfield approach to SQM with N=2 supersymmetries (to confine ourselves
with most simple model). In Section 3 we consider full classical solutions
and the corresponding nilpotent ''marsh''~--- trajectories with nilpotent
addition. In section 4 we investigate the condition of vanishing of half
number of Grassmann constants, because only in this case we obtain exactly
correct number of fermionic zero modes. In section 5 we obtain Witten's
supersymmetric Lagrangian as a consequence of anilpotence of action on such
trajectories. In section 6 we obtained SWKB rules for quantization. In
section 7 we conclude with some comments.

\section{Superfield Formulation}

In $\left( 1|2\right) $ dimensional Euclidean superspace and in the chiral
basis $\left( \tau, \theta ^{+}, \theta ^{-}\right) $ a scalar superfield $%
\Phi \left( \tau, \theta ^{+}, \theta ^{-}\right) $ has the following
expansion
\begin{equation}  \label{1}
\Phi \left( \tau, \theta ^{+}, \theta ^{-}\right) =q\left( \tau \right)
+\theta ^{+}\psi _{-}\left( \tau \right) +\theta ^{-}\psi _{+}\left( \tau
\right) +\theta ^{+}\theta ^{-}F\left( \tau \right)  ,
\end{equation}
where $q\left( \tau \right) $ is a bosonic coordinate, $F\left( \tau \right)
$ is an auxiliary bosonic field, $\psi _{\pm }\left( \tau \right) $ are
fermionic coordinates. Using chiral derivatives $D^{\pm }=\partial /\partial
\theta ^{\mp }+\theta ^{\pm }\partial /\partial \tau $ the superfield
Lagrangian of the model with a superpotential $V\left( \Phi \right) $ can be
written as
\begin{equation}  \label{2}
{\cal L}_E=\frac 12D^{+}\Phi \cdot D^{-}\Phi +V\left( \Phi \right)
\end{equation}
and corresponding equations of motion are\footnote{%
Prime is differentiation by argument.}
\begin{equation}  \label{3}
\frac 12\left[ D^{+}, D^{-}\right] \Phi -V^{\prime }\left( \Phi \right) =0 %
 .
\end{equation}
The Euclidean action has the standard form
\begin{equation}  \label{4}
S_E=\int dtL_E  ,
\end{equation}
where
\begin{equation}  \label{5}
L_E=\int d\theta ^{-}d\theta ^{+}{\cal L}_E
\end{equation}
is an ordinary Euclidean Lagrangian. Expanding (\ref{2}) in $\theta $-series
which is finite due to nilpotence of $\theta ^{\pm }$, using the
nondynamical equation $F=-V^{\prime }\left( q\right) $ and integrating by in
(\ref{5}) $\theta ^{\pm }$ we obtain\footnote{%
Dote is differention by $\tau $.}
\begin{equation}  \label{6}
L_E=\frac{\dot q^2}2+\frac{V^{\prime 2}\left( q\right) }2+\psi _{+}\dot \psi
_{-}+V^{\prime \prime }\left( q\right) \psi _{+}\psi _{-}  .
\end{equation}

\section{Classical Solutions and Even Nilpotent Directions}

From (\ref{6}) we find the following equations of motion
\begin{eqnarray}
\stackrel{..}{q}-V^{\prime }\left( q\right) V^{\prime \prime }\left(
q\right) +V^{\prime \prime \prime }\left( q\right) \psi _{+}\psi _{-} &=&0 %
 ,  \label{7a} \\
\dot{\psi}_{\pm }\mp V^{\prime \prime }\left( q\right) \psi _{\pm } &=&0 %
 .  \label{7b}
\end{eqnarray}
The classical solution of the system (\ref{7a})--(\ref{7b}) is much
simplified if one takes $\psi _{\pm }^{cl}=0$ \cite{coo/fre,sol/hol}. Let us
write a general instanton and antiinstanton solutions without vanishing
fermionic classical coordinates \cite{aku,aku/dup} in the following form
\begin{equation}  \label{8}
\left\{
\begin{array}{c}
q_{+}=q_{0+}- \frac{\lambda _{+}\lambda _{-}}{2V^{\prime }\left(
q_{0+}\right) }, \;\psi _{+}=\lambda _{+}V^{\prime }\left( q_{0+}\right)
,\;\psi _{-}=\frac{\lambda _{-}}{V^{\prime }\left( q_{0+}\right) },\;
{inst}.
\\
q_{-}=q_{0-}+ \frac{\lambda _{+}\lambda _{-}}{2V^{\prime }\left(
q_{0-}\right) },\;\psi _{+}=\frac{\lambda _{+}}{V^{\prime }\left(
q_{0-}\right) },\;\psi _{-}=\lambda _{-}V^{\prime }\left( q_{0-}\right) ,\;{%
antiinst}.
\end{array}
 \right.
\end{equation}
where $\lambda _{\pm }$ are Grassmann integration constants and $q_{0\pm
}\left( \tau \right) $ is a standard instanton/antiinstanton solution \cite
{sol/hol} which can be found from the equations
\begin{equation}  \label{9}
\dot q_{0\pm }\left( \tau \right) =\pm V^{\prime }\left( q_{0\pm }\right) %
 .
\end{equation}
The classical solutions (\ref{8}) can be presented in more compact form
\begin{equation}  \label{10}
q_k=q_{0\left( k\right) }-\frac \nu {2W\left( q_{0\left( k\right) }\right)
},\;\psi _{+\left( k\right) }=\lambda _{+}W^k\left( q_{0\left( k\right)
}\right) ,\;\psi _{-\left( k\right) }=\lambda _{-}W^{-k}\left( q_{0\left(
k\right) }\right)  ,
\end{equation}
where $\nu =\lambda _{+}\lambda _{-}, \nu ^2=0$, $W\left( q\right)
=V^{\prime }\left( q\right) $, $k=+1$ for instantons and $k=-1$ for
antiinstantons. The corresponding scalar superfields (\ref{1}) are
\begin{equation}  \label{11}
\Phi _{\left( k\right) }\left( \tau, \theta ^{+}, \theta ^{-}\right)
=q_{0\left( k\right) }\left( \tau \right) +\theta ^{+}\lambda
_{-}W^{-k}+\theta ^{-}\lambda _{+}W^k+\theta ^{+}\theta ^{-}\left( -W+\nu
\frac{W^{\prime }}{2W}\right)  .
\end{equation}
It is important to stress that despite the fact that the classical
trajectories have a nilpotent part proportional to $\nu $ (\ref{10}) the
action has no such additional parts and it is equal to the action on
instanton trajectories, i.e.
\begin{equation}  \label{12}
S_{Ek}=\int W^2\left( q_{0\left( k\right) }\right) d\tau
=S_{Eq_{0\left( k\right) }}  . \end{equation} For the
topological charge $Q_k$ of instantons/antiinstantons we obtain
\begin{eqnarray}
Q_{k} &=&\frac{1}{2q_{0}\left( \infty \right) }\int\limits_{-\infty
}^{\infty }\dot{q}_{k}^{2}d\tau =  \nonumber \\
&&Q_{0\left( k\right) }+\frac{\nu }{2q_{\left( k\right) }\left( \infty
\right) }\left[ \frac{1}{W\left( q_{\left( k\right) }\left( +\infty \right)
\right) }-\frac{1}{W\left( q_{\left( k\right) }\left( -\infty \right)
\right) }\right]  .  \label{13}
\end{eqnarray}
From (\ref{13}) we conclude that the nilpotent part of the topological
charge vanishes in case of spontaneous symmetry breaking, i.e. when $W\left(
q\right) =V^{\prime }\left( q\right) $ is an even function. That gives
another meaning to the conclusions of \cite{coo/fre,sol/hol}.

\section{Fermionic Zero Modes}

Let us consider a contribution of zero modes in quantum evolution of our
system. In general we have
\begin{equation}  \label{z1}
\Phi _{quant}\left( \tau, \theta ^{+}, \theta ^{-}\right) =\Phi \left( \tau
+\tau _0, \theta ^{+}+\eta ^{+}, \theta ^{-}+\eta ^{-}\right) +\Sigma
^{\prime },
\end{equation}
where $\tau _0, \eta ^{+}, \eta ^{-}$ are shifts corresponding zero modes, $%
\Sigma ^{\prime }$ is contribution of nonzero modes, $\Phi \left( \tau,
\theta ^{+}, \theta ^{-}\right) $ is a classical superfield (\ref{11}).
Expanding the action next to the classical solution or differentiating of
the classical equation of motion by corresponding parameter we find the
equation for zero modes

\[
\{1/2[D^{+}, D^{-}]-V^{\prime\prime}(\Phi )\}\Phi _0=0
\]
But we can find the zero modes differentiating by the parameters of the
solutions (9). We see that the zero modes connected to differentiating of $%
\psi _{-}$by $\lambda _{\_}$and $\psi _{+}$by $\lambda _{+}$ reduce to
inverse functions. But from Cauchy-Schwarz-Bounjakovsky inequality

\begin{center}
$\left( \stackrel{\infty }{\int_0}\psi _{-0}dt\right)
^2\left( \stackrel{%
\infty }{\int_0}\psi _{+0}dt\right) ^2\leq\left( \stackrel{\infty }{\int_0}%
\psi _{-0}\times \psi
_{+0}dt\right) ^2$ \end{center}

where $\psi _{-0}$ and $\psi _{+0}$ are fermionic zero modes, one can see
that only one of two fermionic zero mode can be normalized and consequently
one of two Grassmann constants have to vanish. But a vanished Grassmann
constant corresponds to N=1 broken supersymmetry. Thus we see that one half
of supersymmetries must be broken.

\section{Supersymmetry as Anilpotence of Lagran\-gian}

Let us consider nonsupersymmetric Lagrangian similar to (\ref{6}) as
\begin{equation}  \label{26}
L_{E(nonsusy) }=\frac{\dot{q}^{2}}{2}+\frac{W^{2}\left( q\right) }{2}+
\psi _{+}\dot{\psi}_{-}+U\left( q\right) \psi _{+}\psi _{-}  ,
\end{equation}
where $W\left( q\right) $ and $U\left( q\right) $ are some functions.
Equations of motion are
\begin{eqnarray}
\stackrel{..}{q}-W\left( q\right) W^{\prime }\left( q\right) +U^{\prime
}\psi _{+}\psi _{-} &=&0,  \label{27} \\
\dot{\psi}_{\pm }\mp U\left( q\right) \psi _{\pm } &=&0  .
\label{28}
\end{eqnarray}
Fermionic classical solutions are
\begin{equation}  \label{29}
\psi _{\pm }=\lambda _{\pm }e^{\pm \int U\left( q\right) d\tau }
 , \end{equation} and therefore $\psi _{+}\psi _{-}=\lambda
_{+}\lambda _{-}=\nu =const$ on equations of motion. Let
\begin{equation}  \label{30}
q\left( \tau \right) =q_{0}\left( \tau \right) +\nu q_{N}\left( \tau \right) %
 ,
\end{equation}
where $q_{0}\left( \tau \right) $ is nonsupersymmetric instanton solution $%
\dot{q}_{0}=\pm W\left( q_{0}\right) $ and $q_{N}\left( \tau \right) $
satisfies the following equation
\begin{equation}  \label{30a}
\dot{q}_{N}=W^{\prime }\left( q_{0}\right) q_{N}-\frac{U\left( q_{0}\right)
}{W\left( q_{0}\right) }  .
\end{equation}
Then we obtain for on-shell Lagrangian (\ref{26})
\begin{equation}  \label{31}
L_{E({nonsusy}) }^{n-shell}=W^{2}\left( q_{0}\right) +\nu W\left(
q_{0}\right) \left[ 2W\left( q_{0}\right) W^{\prime }\left( q_{0}\right)
q_{N}-U\left( q_{0}\right) \right]  .
\end{equation}
The requirement of vanishing of the nilpotent part in (\ref{31}) gives
\begin{equation}  \label{32}
q_{N}=\frac{U\left( q_{0}\right) }{2W\left( q_{0}\right) W^{\prime }\left(
q_{0}\right) }  .
\end{equation}
So that from (\ref{30a}) and (\ref{32}) we finally obtain \cite{aku}
\begin{equation}  \label{33}
\frac{U^{\prime }}{U}=\frac{W^{\prime \prime }}{W^{\prime }}
\end{equation}
which can be solved by
\begin{equation}  \label{34}
U=cW^{\prime }  ,
\end{equation}
where $c$ is an even constant which can be taken as 1. Substituting (\ref{34}%
) into $L_{E({nonsusy}) }$ (\ref{26}) gives supersymmetric Lagrangian
$L_{E({susy}) }$ (cf. (\ref{6})) in the following form
\begin{equation}  \label{35}
L_{E({susy}) }=\frac{\dot{q}^{2}}{2}+\frac{W^{2}\left( q\right) }{2}+\psi
_{+}\dot{\psi}_{-}+W^{\prime }\left( q\right) \psi _{+}\psi _{-}  ,
\end{equation}
Thus, the requirement of anilpotence of the on-shell Lagrangian gives us the
supersymmetry condition for the on-shell Lagrangian \cite{aku}.

\section{Quasiclassical Quantization Rules as Anilpotence of Action}

Let us return to Minkowski $\left( 1|2\right) $ superspace by the
substitution $\tau =it$. Then the action instead of (\ref{4}) will
take the form \begin{equation}  \label{14} S=\int L dt  ,
\end{equation}
where\footnote{%
In this section a dote denotes differentiation by $t$.}
\begin{equation}  \label{15}
L=\frac{\dot q^2}2-\frac{V^{\prime 2}\left( q\right) }2+\psi _{+}\dot \psi
_{-}+V^{\prime \prime }\left( q\right) \psi _{+}\psi _{-}  .
\end{equation}
The corresponding equations of motion are
\begin{eqnarray}
\stackrel{..}{q}+V^{\prime }\left( q\right) V^{\prime \prime }\left(
q\right) -V^{\prime \prime \prime }\left( q\right) \psi _{+}\psi _{-} &=&0 %
 ,  \label{16a} \\
\dot{\psi}_{\pm }\pm iV^{\prime \prime }\left( q\right) \psi _{\pm }
&=&0 %
 .  \label{16b}
\end{eqnarray}
Their solutions in case we admit nonzero fermionic classical solutions \cite
{aku/dup} have the following form
\begin{equation}  \label{17}
q=q_0+\nu \frac W{2E},\;\psi _{\pm }=\lambda _{\pm }e^{\mp i\arcsin
\left( \frac W{\sqrt{2E}}\right) }  , \end{equation} where $E$ is
energy of the system and $W=V^{\prime }\left( q_0\right) $. Here
$q_0\left( t\right) $ is a standard solution of 1-dimensional
nonsupersymmetric system which can be found from the equation
\begin{equation}  \label{18}
dt=\frac{q_0}{\sqrt{2E-W^2}}  .
\end{equation}
The corresponding Maupertuis action is
\begin{equation}  \label{19}
S_M=\oint \left( p dq+\psi _{+}d\psi _{-}\right)  ,
\end{equation}
where
\begin{equation}  \label{20}
p=\sqrt{2E-W^2}\left( 1+\nu \frac{W^{\prime }}{2E}\right)
\end{equation}
is a canonical momentum for $q$. On equations of motion the action (\ref{19}%
) takes the form
\begin{equation}  \label{21}
S_M=\oint \sqrt{2E-W^2}dq_0+\nu \oint \left( \frac 1E-\frac
1{2E-W^2}\right) \sqrt{2E-W^2}W^{\prime }dq_0  .
\end{equation}
Second term with nilpotent addition vanishes after integration over a closed
path and we obtain the standard super WKB rules \cite{com/ban/cam}.

The quasiclassical wave function is also defined by classical trajectories
as follows
\begin{equation}  \label{22}
\Psi \left( q,\psi _{+},\psi _{-}\right) \approx A_{\psi }^{\frac i\hbar
S_\psi }  ,
\end{equation}
where
\begin{equation}  \label{23}
A_\psi ={\rm Ber}\left| -\frac{\partial ^2S_{\psi }}{\partial q\left(
0\right) \partial q\left( t\right) }\right|
\end{equation}
is a super generalization of the van Fleck determinant. Using the equations
of motion (\ref{16a})--(\ref{16b}) we obtain
\begin{eqnarray}
A_{\psi } &=&\frac{1}{\sqrt{2E-W^{2}}}\left[ 1+\frac{\nu }{2}\left( \frac{%
W^{\prime }}{2E}+\frac{W^{\prime 2}}{4E^{2}}e^{i\arcsin \left(
\frac{W}{\sqrt{2E}}\right) }\right) \right] ,  \label{24} \\
S_{\psi } &=&\int dt\left( E-W^{2}\right) \left( 1+\nu
\frac{W^{\prime }}{E}%
\right)  .  \label{25} \end{eqnarray}
\newpage
\section*{Conclusions}

Thus we have considered the generalized solutions for the classical
equations of motion for a bosonic-fermionic system which contain the
nilpotent additions~-- ''nilpotent marsh''. In consequence of conservation
the nilpotent constant $\nu $ the particle will ''live'' on such trajectory
and never will have the intersections with usual classical trajectory. But
on such instanton (antiinstanton) trajectories the classical action has a
nilpotent addition, which vanishes only for supersymmetric Lagrangian. The
nilpotent addition to the topological charge vanishes for the superpotential
with spontaneous broken supersymmetry. We studied the normalized zero modes
for the instanton and antiinstanton solutions. As a consequence of
Cauchy-Schwarz-Bounjakovsky inequality for fermionic zero modes only half of
then can be normalized. So we have to vanish half of Grassmann constants,
which corresponds to $N=1$ broken supersymmetry. In contrast of paper \cite
{man0} we have used a superfield approach to SQM \cite{aku/pas}. We have
considered a quasiclassical action on such trajectories and obtained Super
WKB rules for quantization, as an application of our approach.

\section*{Acknowledgments}

We thank A. Pashnev, A. Burinsky, A. Gumenchuk and A. Zheltukhin for
discussions and interest to the work.

This work is partly (A.V.) supported by grants INTAS-93/127ext,
INTAS-93/493ext, INTAS-96/0308.

\end{document}